\documentclass[%
 reprint,
 amsmath,amssymb,
 aps,nofootinbib, longbibliography, floatfix
]{revtex4-1}
\usepackage{bm}
\PassOptionsToPackage{linktocpage}{hyperref}
\usepackage[hyperindex,breaklinks]{hyperref}
\usepackage{enumitem}
\usepackage{slashed}

\renewcommand{\theta}{\vartheta}

\usepackage{array}
\usepackage{mathtools}
\usepackage{xcolor}

\usepackage{etoolbox}
\smallskipamount
\begin{document} 

\title{$SU(5)$ grand unification and $W-$boson mass}

\author{Goran Senjanovi\'c$^{1,2}$ and Michael Zantedeschi$^{1,3}$}
\affiliation{%
$^1$
Arnold Sommerfeld Center, Ludwig-Maximilians University, Munich, Germany
}%
\affiliation{%
$^2$
International Centre for Theoretical Physics, Trieste, Italy
}%
\affiliation{%
$^3$
Max-Planck-Institute for Physics, Munich, Germany
}%

\begin{abstract}
A realistic extension of the minimal $SU(5)$ theory consisting of the addition of an adjoint fermion is known to predict light real fermion and scalar weak triplets, potentially accessible at the LHC. These particles, in addition to playing a key role in gauge coupling unification,
have profound phenomenological implications. The fermion triplet, that through the seesaw mechanism offers a testable origin of neutrino mass, has been already extensively discussed.
 The scalar triplet develops a vacuum expectation value that modifies the $W$-boson mass. 
We show that its low-energy effective theory is remarkably predictive: in the leading approximation, all the relevant physical processes involving the scalar triplet depend only on its mass and the deviation from the Standard Model $W$-mass value.
\end{abstract}
\maketitle

\section{Introduction} 
It is generally accepted that Gran Unification of electroweak and strong interaction is one of the most appealing and well-motivated extensions of the Standard Model (SM). After all, it leads to proton decay and predicts the existence of magnetic monopoles. 

Unfortunately, its minimal realization, the celebrated Georgi-Glashow model~\cite{Georgi:1974sy} based on $SU(5)$ gauge unification is ruled out on two fundamental grounds: it fails to account for the unification gauge couplings and just as the SM, predicts massless neutrinos. 

Some years ago it was shown that a simple realistic extension~\cite{Bajc:2006ia} based on the addition of the adjoint fermion representation, solves at one shot the problems of gauge coupling unification and the vanishing of neutrino masses. 
This in turn predicts both a fermion and a scalar weak triplets at today's energies (for a detailed analysis of the running see Ref.~\cite{Bajc:2007zf,DiLuzio:2013dda}). 
In other words, the structural consistency of the model requires the existence of an oasis close to $M_W$, as we recapitulate below.  

Notice that we are not saying that there could be such an oasis - rather we wish to stress that such light states are a must in this predictive model. 

The fermion triplet allows to test the origin of neutrino mass at hadron colliders, and its physical consequences have been amply discussed~\cite{Bajc:2007zf,Arhrib:2009mz}. The current bound on its mass from LHC~\cite{Novak:2020jju} is roughly $\rm TeV$. As it will be discussed below, due to largeness and degeneracy of its mass, its contribution to $W-$mass deviation is completely negligible.

On the other hand, due to the mixing with the Standard Model Higgs boson, the scalar triplet develops a small vacuum expectation value (VEV) which modifies the $W$-boson mass. This fits with the recent CDF experiment claim \cite{CDF:2022hxs} of $W$-mass deviation from the SM value. 

What about other realistic extensions of the minimal $SU(5)$ theory? We should stress that here we are , completely driven by minimality, in order to have as predictive as possible low-energy effective theory. In this sense, an appealing candidate becomes the addition of a symmetric scalar representation~\cite{Dorsner:2005fq}, leading to a light charged weak scalar triplet responsible for the generation of neutrino mass, which can also modify $W$-mass. The situation is more involved though due to mass splittings, and recently a detailed study of the CDF anomaly was performed in~\cite{Cheng:2022jyi,Heeck:2022fvl}.

It is noteworthy that similarly the minimal $SO(10)$ grand unified theory with small Higgs representations also predicts a neutral scalar triplet to be at low energies~\cite{Preda:2022izo}. However, it moreover allows for additional light scalars - e.g., an additional light weak doublet - whose presence prevents a predictive low energy effective theory of the $W$-mass deviation. For this reason, the analysis of this theory is beyond the scope of the present work. 

If instead one opts for large representations, in the context of $SO(10)$, it is clear that the situation gets much murkier: not only there can be multiple sources for the $W$-mass deviation, but the particle spectrum is unconstrained.

Since low-energy supersymmetry has dominated the field of grand unification, it is fair to ask what happens in the most predictive of such theories, i.e. the minimal supersymmetric $SU(5)$ model. It is a textbook knowledge that this theory provides successful unification of gauge couplings, however, due to the proliferation of states, it is clear that the scale of supersymmetry does not have to be low (it is easy to show that it can be orders of magnitude above the weak scale).
There is more to it though: the mass splittings between the states in the weak multiplets are know to contribute to the $W$-mass deviation at the loop level. So even if one was to stick to low-energy supersymmetry, it would be hard, if possible at all, to know what causes the possible CDF anomaly. 

In this sense, the theory in question here~\cite{Bajc:2006ia} is kind of unique. Not only it requires the light scalar triplet, but as we show below, it is the only source of CDF anomaly, with a predictive low-energy effective theory. 

 The scalar triplet was also introduced by hand~\cite{Ross:1975fq}, and its phenomenology studied over the years, see e.g., \cite{Gunion:1989ci,Blank:1997qa,Chen:2006pb,FileviezPerez:2008bj}. Recently, it was pointed out that it could explain the CDF result see e.g.~\cite{Strumia:2022qkt,Perez:2022uil}.
However, grand unified theories (GUTs), besides providing the rationale for its existence and its accessibility at today's energies, also makes the low-energy effective theory of the triplet completely predictive: In the leading approximation, all the relevant physical processes of new particles depend only on 
a single dimensionless parameter, the ratio of the possible $W$-mass deviation over its mass. The CDF measurement, if confirmed, would fix this parameter, making the theory  testable at hadron colliders. 

Strictly speaking, there could be a deviation from this result in the neutral sector. This would imply the presence of yet another light particle -- the singlet responsible for the breaking at the grand unified scale\footnote{We thank Selina Kettner for reminding us of the possible fine-tuning of the singlet mass.} ($M_G$) -- of degenerate mass with the triplet, which would be phenomenologically interesting in itself. 

Without the GUT framework, some of the triplet couplings happen to be arbitrary, to the extent that its neutral component could even be artificially decoupled~\cite{Perez:2022uil} from SM fermions. In our case, these couplings are uniquely determined and related to the couplings with gauge bosons.
We show here how this remarkable result comes about.  

\section{${SU(5)}$ and the oasis} 
It is well-known that the three gauge couplings of the SM  do not unify at a single point: $\alpha_1$ meets $\alpha_2$ too early, at about $10^{13}\rm GeV$, while $\alpha_2$ and $\alpha_3$ meet at around $10^{17}\rm GeV$. 

In order to keep proton stable enough, the way out of this impasse is to increase the $\alpha_1-\alpha_2$ meeting point. Since there are no new gauge bosons that could slow down the rise of $\alpha_1$, the only option is to slow down $\alpha_2$ - through new fermions or/and scalars with small (ideally zero) hypercharge. 
 Clearly, the real $SU(2)_L$ scalar triple $T_S$ does the needed job, since it carries zero hypecharge.  

However, even when lying close to the electro-weak scale, this particle does not suffice to ensure gauge coupling unification. On top of that, the minimal model \cite{Georgi:1974sy} predicts a massless neutrino. More is needed. 

A simple addition is the adjoint fermion representation $24_F$ \cite{Bajc:2006ia} which contains a real weak triplet fermion $F$, leading to the following particle content
\begin{equation}
    24_{H};\quad 24_F;\quad 5_H;\quad \overline{5}_F^a; \quad 10_F^a,
\end{equation}
where $a=1,2,3$ denotes the generation index. As well known, compatibility with the SM fermion spectrum requires the presence of higher dimensional operators. 

The scalar adjoint representation $24_H$ provides the GUT symmetry breaking to the SM and that's where the scalar triplet $T_S$ resides, while the triplet $T_F$ belongs to the fermion adjoint representation $24_F$.
These triplet states allow for the successful unification of gauge couplings as can be seen from the unification condition \cite{Bajc:2006ia}
\begin{equation}
    \frac{M_G}{M_W}= \exp{\left[\frac{15\pi}{42}(\alpha_1^{-1}-\alpha_2^{-1})_{M_{W}}\right]\left(\frac{M_W^5}{m_{T_S} m_{T_F}^4}\right)^{\frac{5}{84}}}.
\end{equation}
Proton stability $\tau_p\gtrsim 10^{34}\rm yrs$ \cite{Super-Kamiokande:2020wjk} requires $M_G \gtrsim 10^{15}\rm GeV$, and thus both $T_S$ and $T_F$ need to lie close to present-day energies.
More precisely, if the fermion triplet is to be accessible to the LHC energies, the scalar triplet could be as heavy as $10^3 \rm TeV$\footnote{We are grateful to Borut Bajc for pointing this to us and correcting an error in the original version of this work.}. However, as we show later, the CDF considerations would require $T_S$ to lie below roughly $10\rm TeV$. 

In the rest of this work, we focus on the scalar triplet $T_S$, whose physics has not yet been fully explored. First, we turn our attention to the Higgs potential, with relevant terms 
\begin{equation}
    V_{\rm relevant} = m^2 5_H^{\dagger}5_H + M 5_H^\dagger 24_H 5_H + \beta 5_H^{\dagger} 24_H^2 5_H.
\end{equation}
 To start with, the masses of the SM Higgs doublet $\Phi$ and the triplet $T_S$ must  vanish at this stage. The colored triplet $T_C$ from $5_H$, though, must lie close to the GUT scale since it mediates proton decay. This assumes no judicious cancellations of Yukawa coupling matrices, in which case $T_C$ could lie even at TeV energies~\cite{Dvali:1992hc,DSS}. 

We should stress that the lightness of $T_S$ can be easily achieved~\cite{Guth:1981uk}. 
Moreover, $T_S$ has cubic coupling $\mu \Phi^\dagger T_S \Phi$ that must be also kept small in order to preserve the stability of the Higgs picture of SM symmetry breaking. We show now how this works by omitting higher-dimensional operators in the potential (including them only facilitate things). 

Upon breaking of $SU(5)$ gauge symmetry with $\langle 24_H\rangle = v_G\, {\rm diag}\left(1,1,1,-3/2,-3/2 \right)$, the relevant mass terms become
\begin{equation}
    \begin{split}
       &m_{\Phi}^2 = m^2 - \frac{3}{2}M v_G + \frac{9}{4}\beta v_G^2\\
        &m_{T_C}^2 = m^2 + M v_{G} + \beta v_G^2\\
        &\mu = M - 3\beta v_G.
    \end{split}
\end{equation}
Requiring $m_\Phi\simeq\mu\simeq0$ leads to $m_{T_C}^2=25 \beta v_G^2/4$, implying $\beta\gtrsim 10^{-3}$ for the sake of proton's stability.

In short, as in the minimal $SU(5)$ theory, the desired spectrum is achieved by fine tuning the couplings in the potential.

\section{Low energy effective theory} 
As well known \cite{Buras:1977yy}, $T_S=T$ attains a non-vanishing VEV $v_T$ due to the tadpole interaction with $\Phi$. This can be seen from the Lagrangian
\begin{equation}
\begin{split}
\label{eq:lagrangian}
    &\mathcal{L} = |D_\mu \Phi|^2 +{\rm Tr} (D_\mu T)^2  +m_\Phi^2 \Phi^\dagger \Phi- m_{T}^2{\rm Tr}\,T^2\\ &-\lambda_T \left(\rm{Tr}\,T^2\right)^2  -\lambda_\Phi (\Phi^\dagger \Phi)^2 - \rho \Phi^\dagger \Phi {\rm Tr}\,T^2 - \mu \Phi^\dagger T \Phi,
    \end{split}
\end{equation}
where $T=T^{\dagger}$ transforms as $T\rightarrow U T U^\dagger$, with covariant derivative $D_\mu T = \partial_\mu T + i g \left[A_\mu,T \right]$. $T$ and $\Phi$ are decomposed as
\begin{equation}
\label{eq:deltadec}
    \Phi = 
    \frac{1}{\sqrt{2}}\begin{pmatrix}
    \sqrt{2}\phi^+\\
v+\phi_R^0 + i \phi_I^0 
    \end{pmatrix},\,
    T = \frac{1}{2}
    \begin{bmatrix}
    v_T + t^0 &\sqrt{2}t^+\\
\sqrt{2}t^- &-v_T - t^0
    \end{bmatrix}.
\end{equation}

Upon $SU(2)$ breaking, from the $\mu$-term in \eqref{eq:lagrangian},
$v_T$ is readily computed
\begin{equation}
\label{eq:vdelta}
    v_{T}\simeq  \frac{\mu}{g^2} \left(\frac{M_W}{m_T}\right)^2,
\end{equation}
where we used the tree-level SM mass relation $M_W^{SM} \simeq g v/2$.
In the above expression, $m_T$ stands for the triplet mass, corrected by the $\rho$ term. For simplicity, we use the same symbol as in \eqref{eq:lagrangian}. Eq.~\eqref{eq:vdelta} holds true if $v_T\ll M_W$, compatible with  the high-precision success of the SM. 

This is an example of the hierarchy of weak isospin breaking. In theories with a large mass scale $M \gg M_W$, higher isospin fields have tadpole induced VEVs proportional to $M_W^{n}/M^{n-1} $. The integer $n$ counts the minimal number of SM doublets that couple to the  field in question in order to form a SM invariant expression. This can be understood as the decoupling of heavy SM singlet fields: when $M\rightarrow \infty$, all physical effects must disappear. In the case of the SM Higgs, $n=1$ implying $v\sim M_W$. For our triplet $n=2$ (see the last term in \eqref{eq:lagrangian}) and \eqref{eq:vdelta} is recovered in the limit $\mu \simeq m_{T} = M$. Clearly, larger weak isospin fields have more suppressed VEVs as can be shown by direct inspection. 

The mass spectrum of the triplet then becomes
\begin{equation}
\label{eq:massestriplet}
    \begin{split}
    m_T^2 \simeq m_{t^{\pm}}^2 = m_{t^0}^2 +\mathcal{O}(v_T^2),
    \end{split}
\end{equation}
and is independent of the sign in front of $m_T^2$ in \eqref{eq:lagrangian}. In turn, $m_{t^{\pm}}\gtrsim 250\rm GeV$ \cite{Chiang:2020rcv}, forces the same bound on $m_T$ since the last contribution in \eqref{eq:massestriplet} is negligible. Thus, whether or not the triplet is Higgsed\footnote{We are grateful to G. Dvali for helping us clarify this essential point.}, the mass spectrum is dominated by $\mu$ and is therefore degenerate. 
Radiative corrections further split the masses, but are negligible \cite{Cirelli:2005uq} for the present discussion. It should be stressed that, for the same reason, the fermionic triplet cannot account for the CDF anomaly. First of all, the lower limit on its mass is even bigger, around $\rm TeV$ as commented in the introduction, and moreover, the mass splitting results only at the loop level and is thus negligible.

What about the upper limit on $m_T$? It is easy to demonstrate that the stability of the SM vacuum leads to the following result 
\begin{equation}
\label{eq:upperlimit}
    m_T,\mu \lesssim \frac{v^2}{v_T}\,.
\end{equation}
Clearly, a vanishing $v_T$ would allow for arbitrarily large mass scales $m_T$ and $\mu$ - it is only the unification consistency that forces the triplet to be light.

Once $v_T$ is turned on, the $\mu$-term leads to the physically important mixing $\theta$ between $\Phi$ and $T$. 

For the charged components, the physical states, in the small mixing approximation are
\begin{equation}
\label{eq:physicalplus}
\begin{split}
    &G^+ \simeq \phi^+ - \theta t^+, \\
    &H^+ \simeq t^+ + \theta \phi^+.
\end{split}
\end{equation}
where $G^+$ is the would-be Goldstone boson eaten by $W^+$, and $H^+$ is a heavy charged particle mostly from the triplet. The small mixing angle $\theta$ is given by the properly normalized ratio of the two VEVs
\begin{equation}
\label{eq:theta}
   \theta =  \frac{g\, v_T}{ M_W},
\end{equation}
where $\mu$ was traded for $v_T$ using \eqref{eq:vdelta}.
In the neutral sector, instead, the physical states become
\begin{equation}
\label{eq:physical}
\begin{split}
    &h^0 \simeq \phi^0 + \theta \left(\frac{m_H^2}{m_H^2-m_h^2} \right) t^0 \\
    &H^0 \simeq t^0 - \theta\left(\frac{m_H^2}{m_H^2-m_h^2} \right) \phi^0,
    \end{split}
\end{equation}
where $h^0$ and $H^0$ stand for the SM Higgs boson and the neutral component of the
triplet, respectively. In \eqref{eq:physical}, $m_h= m_\Phi$ and $m_H = m_T$ are the masses of the physical states. Moreover, the effect of the $\rho$-interaction in \eqref{eq:lagrangian} on $\theta$ has been neglected (such correction is not affecting the charged sector). This is because, as we will show below, UV consistency of the theory requires $\rho\ll1$, which in turn leads to $\rho v_T\ll \mu/2$ (notice that $\mu\gtrsim 20 \rm GeV$ from \eqref{eq:vdelta} and $v_T\lesssim 5\rm GeV$ from electroweak precision data).
 
Eqs.~\eqref{eq:physicalplus} and \eqref{eq:physical} are given in the physical limit of small $\theta$, justified by $v_T \ll M_W$. 
 Apart from a small phenomenological region for $m_H\gtrsim 250\rm GeV$, corresponding to its lower bound, the mixings in \eqref{eq:physical} and \eqref{eq:physicalplus} are, up to a sign, identical. In what follows, we will give analytical formulas relevant for $m_H\gtrsim500{\rm GeV}$. The small corrections in the lower end of the spectrum will be shown in the Figs.~\ref{fig:decays}-\ref{fig:decaysplus} below.
 
Although not predicted to be light, the mass of the singlet $S$ in $24_H$ could be, in principle, fine-tuned to be much smaller that the GUT scale. In this case, $S$ and $H^0$ mix through the electro-weak symmetry breaking. Unless $S$ and $H^0$ are taken to be degenerate in mass, their mixing $\alpha$ is small due to the large mass of $H^0$ and can be safely ignored in~\eqref{eq:physical}. In the case of approximate degeneracy, however, the mixing in \eqref{eq:physical} is arbitrary and no longer determined by \eqref{eq:theta}.
Then, however, one has a new prediction of yet another state with the mass close to the mass of the triplet. We will ignore this possibility in the following.

As commented above, small $\theta$ gets corrected by the $\rho$ term in the neutral sector. However, this contribution is subleading due to the requirement that the theory stays perturbative all the way to the unification scale. This can be seen from the 1-loop renormalization group flow \cite{Cheng:1973nv} of the quartic couplings in \eqref{eq:lagrangian}
\begin{equation}
\label{eq:RG}
\begin{split}
	&\frac{d\lambda_\Phi}{dt}\simeq \left(\frac{d\lambda_\Phi}{dt}\right)_{SM}+ \frac{2}{3\pi^2}\rho^2,
   \\
&  \frac{d\lambda_T}{dt}\simeq\frac{1}{16\pi^2}\left( {24 {g_2}^4-24 {g_2}^2 \lambda_T+11 \lambda_T^2+16 \rho ^2}\right),\\
&\frac{d\rho}{dt} \simeq\frac{1}{32\pi^2}\left( 12 \rho 
  y_t^2-3 {g_1}^2 \rho +6
   {g_2}^4-33 {g_2}^2 \rho\right.\\ &\,\,\quad\qquad\qquad\qquad\qquad\left.+10 \lambda_T  \rho +24 {\lambda_{\Phi}} \rho
   +16 \rho ^2\right),
 \end{split}
 \end{equation}
 where $t = \log \mu$ and the first term on the right-hand side of the first equation denotes the SM running of the quartic Higgs coupling. We neglect small Yukawa couplings - including the ones of $T_F$, which must be tiny in order to guarantee the smallness of neutrino masses. 
 Eqs.~\ref{eq:RG}, together with the gauge couplings and top Yukawa coupling renormalization group equations were then solved numerically between $M_W$ and $M_G$. 
 
 First of all, $\rho$ increases the value of $\lambda_\Phi$ at $M_G$, which, in the SM, is known to be negative at high energies \cite{Degrassi:2012ry}. While the inclusion of uncertainties in the SM input parameters combined with a higher-loop analysis seem to somehow alleviate the issue \cite{Degrassi:2012ry} - bringing the coupling close to zero - it is noteworthy that the triplet can ease this tension, as seen from the first equation in \eqref{eq:RG}. 

In short, the central value of the quartic Higgs coupling turns negative at around $10^{10}\rm GeV$ if no new particle contributes to the running. This poses a challenge for the desert picture, providing yet another rationale for the oasis near $M_{\rm W}$.

\begin{figure}[t]
    \centering
    \includegraphics[width=0.46\textwidth]{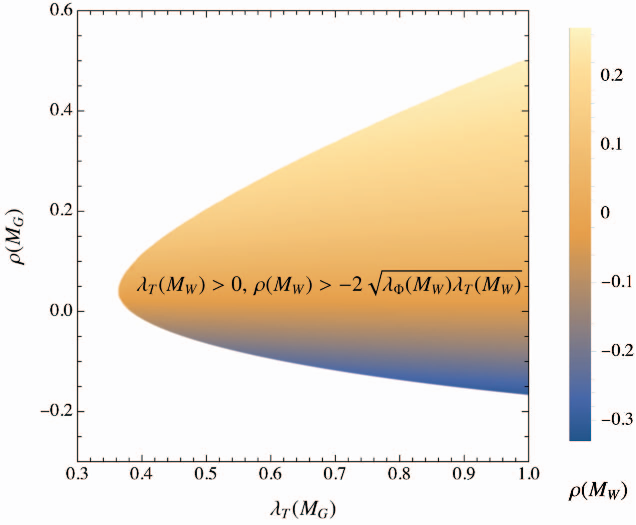}
    \caption{Allowed region from the 1-loop renormalization group flow. Vacuum stability conditions at $M_{\rm W}$ are shown in the plot.}
    \label{fig:rg}
\end{figure}

As seen in Fig.~\ref{fig:rg}, the requirements of perturbativity at $M_G$ and vacuum stability at $M_W$ imply $|\rho(M_W)|\lesssim 0.3$.
This translates into a correction to $\theta$ in \eqref{eq:physical} which is negligible for the triplet mass above $500$ GeV, as shown in Fig.~\ref{fig:decays} below. The rotation to the physical basis in the charged sector \eqref{eq:physicalplus} is unaffected by $\rho$-value.

Finally, the $\lambda_T$ term is negligible for tiny $v_T$. The end result is striking. The apparently complicated structure in \eqref{eq:lagrangian}, gets tremendously simplified: The only physical parameters are $\theta$ and $m_H$. 

The first and foremost impact of $v_T$, or better $\theta$, is the modification of the $W$-mass
\begin{equation}
\label{eq:mwmass}
    M_{W}^2\simeq (M_W^{SM})^2(1+\theta^2), \quad M_Z^{2}= (M_Z^{SM})^2,
\end{equation}
where tiny loop effects from both $T_F$ and $T_S$ are neglected due to the degeneracy of their mass spectra.
Note that the $Z$-mass is unchanged since $T$ carries no hypercharge. 
If one accepts the CDF result $M_W=80.433 \pm 0.009$ \cite{CDF:2022hxs}, by taking $M_W^{SM}=80.357 \pm 0.006\rm GeV$ \cite{Zyla:2020zbs}, for $g(M_{W})= 0.6517$ \cite{Zyla:2020zbs}, it follows from \eqref{eq:mwmass} that $\theta\simeq 0.04$. Moreover, as it can be seen from (\ref{eq:upperlimit}), this would also force $T_S$ to lie below $10\rm TeV$, in order to preserve the Higgs picture of the SM. One may be surprised that a $\rm TeV$ particle is not decoupling from the SM physics. However it must be kept in mind that the impact is direct through the tree level VEV and $W-$mass deviation is tiny.

We close this sections with the relevant couplings of $H^0$ and $H^+$ to the SM particles. Ignoring the conspiracy of the  singlet mass tuned not only much below $M_G$, but to be degenerate with $m_T$, one gets
\begin{equation}
\label{eq:couplings}
\begin{split}
   &g \theta  H^0:\\
   &\quad \,\,\, M_W  WW , \,\frac{M_Z Z^2}{2c_W} ,\, \frac{(m_H h^0)^2}{4 M_W} , \frac{m_f\overline{f}f}{2M_W},\\
  &g\theta H^+: \\
  &\quad \,\,\, \frac{M_W ZW^-}{c_W} ,\, \frac{i}{2}\overset{\leftrightarrow}{\partial} h^0 W^- , \frac{m_d\overline{u}_L d_R - m_u \overline{u}_Rd_L}{\sqrt{2}M_W}.
  \end{split}
\end{equation}
valid in the range $500{\rm GeV} \lesssim m_H \lesssim 10\rm TeV$, where the mixings in the neutral and charged sectors are basically equal.

In case of degeneracy, one would end up with two states whose couplings would depend additionally on the $\alpha$ mixing.

\section{Phenomenological implications} 
The triplet $T$, as given in \eqref{eq:deltadec}, contains a real CP-even field $H^0$ and the charged field $H^+$. They can be produced through the $W$ and $Z$ fusion, both singly (suppressed by the small $v_T$) and pairwise, and $H^+$ also through the photon and the $Z$-boson exchange (for detailed studies see \cite{Gunion:1989ci,Blank:1997qa,Chen:2006pb,FileviezPerez:2008bj,Chiang:2020rcv}). For explicit searches in the gauge channels at LHC see, e.g.,~\cite{CMS:2019uys,CMS:2021wlt}.
Here we study their decays to SM particles in the physical limit $m_{H}\gg M_W$ and $\theta \ll 1$.

\paragraph*{Neutral Component} Once produced, it decays into  pairs of $W$, $Z$ and Higgs bosons, with the rates computed from \eqref{eq:couplings}
\begin{equation}
\begin{split}
\label{eq:neutralbos}
    {\frac{1}{2}} \Gamma(H^0\rightarrow &W^+W^-)\simeq \Gamma(H^0\rightarrow ZZ)\\&\simeq\Gamma(H^0\rightarrow h^0h^0) \simeq\theta^2\frac{g^2}{128 \pi} \frac{m_H^3 }{M_W^2}.
    \qquad
    \end{split}
\end{equation}
The monochromatic final states, especially $W$ and $Z$-bosons, provide clean signature for the $H^0$ detection. Moreover, the width of $H^0$ is narrow; for $m_H \simeq 10 {\rm TeV}$ and $\theta = 0.04$, $\Gamma_H/M_H \simeq 10^{-2}$, similar to the $W$ and $Z$-bosons decay widths. 

It is interesting that the gauge bosons and Higgs final state rates are equal, as expected in the high-energy limit $E \gg M_W$, when the SM symmetry breaking becomes negligible. 

Next, using \eqref{eq:couplings}, the decay rate of $H^0$ into fermion-antifermion pair is given by
\begin{equation}
\label{eq:fermiondecay}
    \Gamma(H^0\rightarrow f\overline f)\simeq {N_c}\theta^2 \frac{g^2}{{32}\pi}\frac{m_f^2m_H}{M_W^2},  
\end{equation}
where $N_c$ denotes the number of colours of the decay product. 
\begin{figure}[t]
    \centering
    \includegraphics[width=0.46\textwidth]{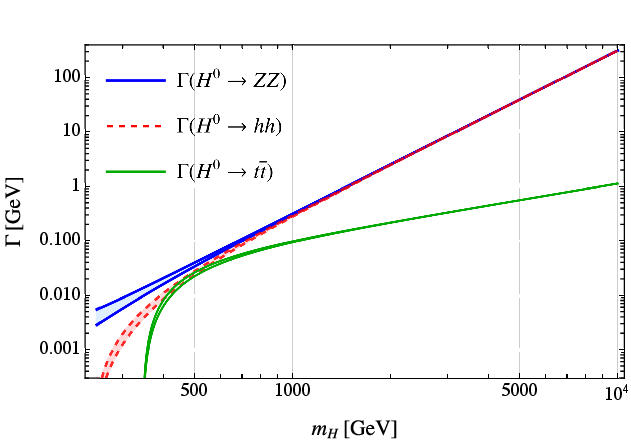}
    \caption{Decay rates of neutral component for $H^0$, $\theta\simeq0.04$.}
    \label{fig:decays}
\end{figure}

In Fig.~\ref{fig:decays}, the full decay rates of $\Gamma(H^0\rightarrow ZZ)$, $\Gamma(H^0\rightarrow h^0h^0)$ and $\Gamma(H^0\rightarrow t\overline{t})$ are shown as a function of $m_H$. The filled area indicates the uncertainty in $\theta$ due to the presence of $\rho$-term in the potential  of \eqref{eq:lagrangian}. As it can be seen, already for $m_H\gtrsim 500\rm GeV$, this area collapses into a line.

Clearly, the fermion modes are negligible, except for the top quark, in which case \eqref{eq:fermiondecay} is comparable to \eqref{eq:neutralbos} for a small region around $m_H\simeq 500 \rm GeV$. As noted above, the $hh$ and $ZZ$ decay rates become practically identical for $m_H\gtrsim 1\rm TeV$.

Since the SM Higgs boson decays into two photons, so does $H^0$, with the rate  suppressed by $\theta$
\begin{equation}
\begin{split}
   \Gamma(H^0\rightarrow \gamma \gamma)\sim \theta^2  \frac{m_H}{M_W}\, \Gamma(h^0\rightarrow \gamma \gamma).
\end{split}
\end{equation}
This expression is just a rough dimensional estimate. 
In the SM, as we know, $\gamma\gamma$ channel was the path towards the Higgs boson discovery: the small rate was compensated by the clear signature of two monochromatic photons, and by the fact that the SM Higgs cannot decay into two on shell $W$-bosons. In this case, the effect is less important since the on-shell $W$ and $Z$ final states decay channels are now allowed and enhanced by large $m_H$. Still, it is a clean channel, worth keeping in mind.

\paragraph*{Charged Component} The charged component $H^+$ decays mainly into a $WZ$ and $Wh$, with the following respective decay rates computed from \eqref{eq:couplings}
\begin{equation}
\begin{split}
     \Gamma(H^+\rightarrow W^+Z)&\simeq \Gamma(H^+\rightarrow W^+h^0)\\&\simeq\Gamma(H^0 \rightarrow W^+W^-).
\end{split}
\end{equation}

\begin{figure}[t]
    \centering
    \includegraphics[width=0.46\textwidth]{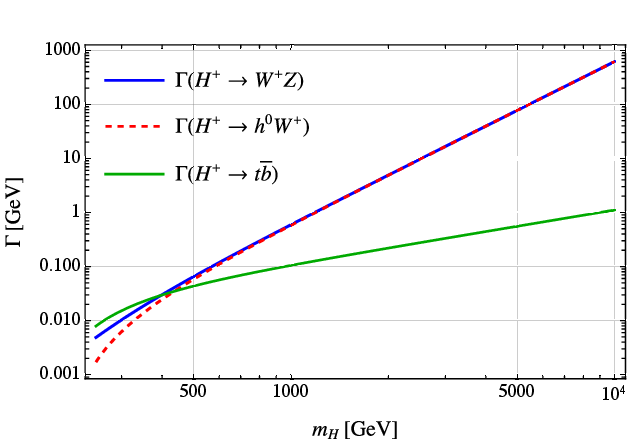}
    \caption{Decay rates of charged component $H^+$, $\theta\simeq0.04$.}
    \label{fig:decaysplus}
\end{figure}

In analogy with \eqref{eq:fermiondecay}, the fermionic decay rates of $H^+$ are negligible, except for $\Gamma(H^+\rightarrow t\overline b)$ which can be computed from \eqref{eq:couplings} and gives
\begin{equation}
    \Gamma(H^+\rightarrow t\overline b)\simeq  \Gamma(H^0\rightarrow t\overline t).
\end{equation}
Notice that since $H_0$ and $H^+$ are degenerate, all the decay rates are correlated. 
This can be seen in Fig.~\ref{fig:decaysplus}. The situation is analogous to the neutral scalar case, except that, for $m_H \lesssim 500\rm GeV$, the fermionic decay dominates, since the kinematics helps (bottom quark is almost massless at these energies). 
In this mass range this is the leading decay channel which is searched for at LHC~\cite{Chen:2022zsh}. 
For higher mass values, once again, the decays into bosons take over.

\paragraph*{Standard Model Higgs} Due to the mixing between the doublet and the triplet, the usual Higgs boson couplings are modified by $\theta^2$, implying a small deviation from its couplings to gauge bosons and fermions, on the order of a permil. While difficult to measure, especially at the hadron colliders, this would eventually serve as an important independent test of the theory.

\section{Summary} 
In this work we have revisited a well-motivated extension of the minimal $SU(5)$ grand unified theory, which predicts a light fermion and a light scalar real weak triplet~\cite{Bajc:2006ia}.
The fermion triplet plays a key-role in generating a small neutrino mass and its physics has been studied extensively in the past. Suffice it to say here that $m_{T_F}\lesssim 10\rm TeV$. 

Here we focus on its scalar counterpart, which naturally develops a small expectation value, implying a deviation of $W$-mass from its SM value. 
The emerging low-energy effective theory is then predictive: there are a  plethora of calculable physical processes which depend only on the triplet mass and a single parameter $\theta$ that measures the $W$-mass deviation. 

In view of the recent CDF announcement \cite{CDF:2022hxs} of the $W$-mass disagreement with its SM value, this theory becomes of phenomenological relevance. If true, the CDF result would fix $\theta$, and implies $m_H\lesssim 10\rm TeV$, making the theory testable at hadron colliders. For a triplet mass above TeV, this becomes a high-precision theory, almost at the level of the SM. 

What would be the implication of the theory if CDF result turned out to be wrong? Although the triplet is generally predicted to be light, its VEV $v_T$ depends on both its mass and the unknown parameter $\mu$ and thus, it can be as small as needed.  In other words, the consistency of the SM does not affect the consistency of the original theory.

In short, a simple and well-motivated grand unified theory predicts new light particle states.
The end result is profound: the theory makes the case for the $W$-mass deviation and ties it to a number of calculable physical processes. The far reaching testable consequences may have just started to unravel themselves.

{\bf Note added} As we were about to post this work, 
a new paper \cite{Evans:2022dgq} appeared in which a particular $SU(5)$ model - very different from the theory presented here - is built in order to accommodate the CDF-anomaly.

{\bf Acknowledgments} We are grateful to Damir Lelas and Gia Dvali for illuminating discussions, and to Damir Lelas and Alejandra Melfo for important comments and indispensable help in the preparation of this manuscript.

\bibliography{biblio}

\end{document}